# Features of the Domain Boundaries of a Highly Anisotropic ($S = 1$) Antiferromagnet near the Transition to the Quantum Paramagnet Phase

V. V. Konev[a], *, V. A. Ulitko[a], D. N. Yasinskaya[a], Y. D. Panov[a], and A. S. Moskvin[a]

[a]*Ural Federal University, Yekaterinburg, 620002 Russia*
*\*e-mail: vitaliy.konev@urfu.ru*

**Abstract**—It is shown that the structure of antiphase domain boundaries in the antiferromagnetic (AFM) phase of a highly anisotropic magnet with $S = 1$ on a two-dimensional square lattice depends greatly on single-ion anisotropy parameter $D$. Computer modeling on large square lattices illustrates the changes in the boundary structure from the quantum paramagnet (QP) to the $XY$ phase, including the intermediate QP–$XY$ phase at fairly small variations in positive $D$.

## INTRODUCTION

In contrast to quantum magnets with $S = 1/2$ spin, systems with $S = 1$ spin are characterized by more complex Hamiltonian, single-ion anisotropy, biquadratic intercentric interactions, and totally new phase states of the quantum paramagnet (QP) type corresponding to an easy-plane phase in the classical approach. The interest in these systems is due to both highly anisotropic magnets based on Ni$^{2+}$ ($S = 1$) (e.g., Y$_2$BaNiO$_5$ [YBNO], Ni(C$_2$H$_8$N$_2$)$_2$NO$_2$(ClO$_4$) [NENP]) [1] and the so-called pseudo-spin systems of the semi–hard core boson type with constraints on filling lattice sites $n = (0, 1, 2)$, or mixed valence ion systems of the triplet type: Cu$^{(1+, 2+, 3+)}$ in cuprates La$_{(2-x)}$Sr$_x$CuO$_4$ and Bi$^{(3+, 4+, 5+)}$ in bismuthates [2, 3]. In all cases, the phase diagrams of spin or pseudo-spin systems with $S = 1$ is considerably richer than those of similar systems with $S = 1/2$ quantum (pseudo)spin, due primarily to the emergence in the Hamiltonian of addends of the single-ion anisotropy and biquadratic interaction types, plus ones of the quantum paramagnet and spin-nematic phase types.

## MODEL

Let bus consider a model cuprate that is a 2$D$ system of Cu centers in a CuO$_2$ plane of cuprates that can be in three different valence charge states: Cu$^{(1+, 2+, 3+)}$. We associate this charge triplet with three states of $S = 1$ pseudo-spin as Cu$^{1+} \to$ M$_S = -1$, Cu$^{2+} \to$ M$_S = 0$, Cu$^{3+} \to$ M$_S = 1$, and use the familiar ways of describing spin systems. The spin algebra of systems with $S = 1$ (M$_S = 0, \pm1$) includes eight independent nontrivial (three dipole and five quadrupole) functionals: $S_z$; $S_\pm = \pm(S_x \pm iS_y)$; $S_z^2$; $T_\pm = \{S_z, S_\pm\} = S_zS_\pm + S_\pm S_z$; and $S_\pm^2$. Incremental/decremental functionals $S_\pm$ and $T_\pm$ change the (pseudo)spin projection to $\pm 1$, but in different ways: $\langle 0|S_\pm|\mp 1\rangle = \langle \pm 1|S_\pm|0\rangle = \mp\sqrt{1}$, and $\langle 0|T_\pm|\mp 1\rangle = -\langle \pm 1|T_\pm|0\rangle = +1$. Incremental/decremental functionals $S_\pm^2$ describe transitions $|-1\rangle \to |+1\rangle$; i.e., they generate on a site either a hole $(S_+^2)$ or an electron $(S_-^2)$ pair that is a composite local boson with kinematic constraint $S_\pm^2 = 0$, emphasizing its nature as a hard-core boson.

Local (on-site) nondiagonal parameter $XY$ of the order of $\langle S_\pm^2 \rangle$, which is actually a parameter of the local superconducting order, is nonzero only when the site hosts a quantum superposition of states $|-1\rangle$ and $|+1\rangle$.

We write the effective Hamiltonian that commutates with the $z$-component of the total spin $n = \frac{1}{N}\sum_i S_{iz}$ and thus maintains the system's magnetization as the sum of potential and kinetic energies: $H = H_{\text{pot}} + H_{\text{kin}}$:

$$H_{\text{pot}} = D\sum_i S_{iz}^2 + J\sum_{\langle ij \rangle} S_{iz}S_{jz}. \qquad (1)$$

In calculating the kinetic energy, we consider only the contribution from double-ion biquadratic anisotropy $H_{\text{kin}} = -t\sum_{\langle ij \rangle}\left(S_{i+}^2 S_{j-}^2 + S_{j+}^2 S_{i-}^2\right)$. The first term in (1) (i.e., the single-ion anisotropy) describes the density–density correlation effects on the sites, while the second term describes inter-site interactions (correlations) of the density–density type. Below, we consider only the interactions between nearest neighbors with positive (antiferromagnetic) signs of inter-center correlation parameter *J*.

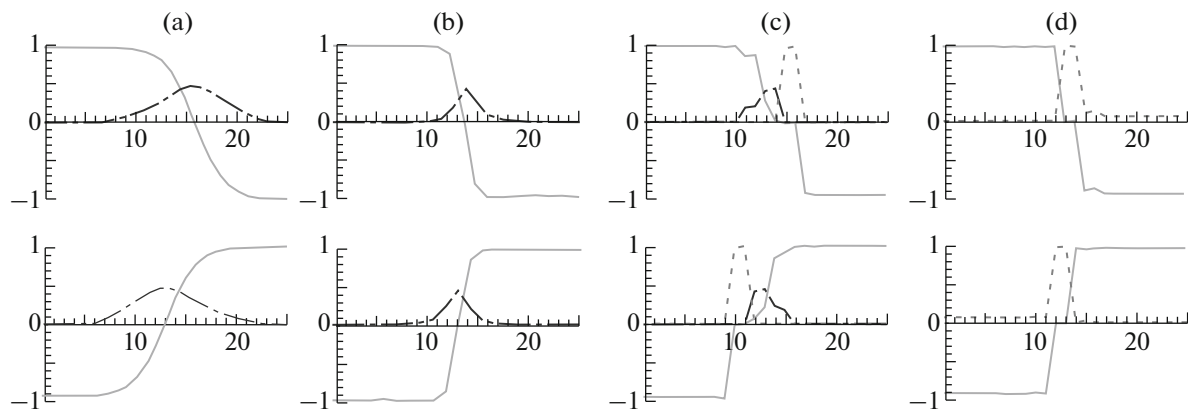

**Fig. 1.** Average distribution across the domain boundary at local parameters on the order of AFM, *XY*, and QP phases marked by solid, dashed-and-dotted, and dashed lines, respectively, on two sublattices *A* and *B* (the top and bottom parts of the figure, respectively) at different values of parameter *D*: (a) −5.0, (b) 1.0, (c) 1.1, and (d) 1.2. The values along the horizontal axis are presented in terms of the lattice constant.

Depending on the relationship between the parameters of Hamiltonian (1) and magnetization (*n*), the system ground state corresponds either to the homogeneous phase of the quantum paramagnet type with $\langle S_z \rangle = \langle S_z^2 \rangle = 0$, which is attained at high positive values of parameter *D* (a large *D* phase); or to the antiferromagnetic (AFM) phase along the *z*-axis; or to the *XY* phase with a nonzero parameter on the order of $\langle S_\pm^2 \rangle$.

## RESULTS AND DISCUSSION

We used an NVidia graphical processing unit for the Monte Carlo modeling of the antiferromagnet phase transition of highly anisotropic magnet *S* = 1 in the two-sublattice approximation on a square lattice of 256 × 256 with periodic boundary conditions at selected parameters *t* =1, *J* = 0.75, *n* = 0.04, which ensured a ground state of the antiferromagnet ordering type in a rather wide range of variations of single-ion anisotropy parameter *D*.

At *D* = −5, a stripe domain structure formed during rapid thermalization (annealing). At low temperatures, a strongly pronounced filamentary *XY* phase emerged at the center of the antiphase domain boundaries of the AFM phase, which was characterized primarily by a nonzero module of the local parameter of the order *XY*. Upon an increase in double-ionic biquadratic anisotropy *t*, the domain boundary gradually broadened and the volume of the *XY* state grew up to the total displacement of the AFM phase and the transition to the inhomogeneous *XY* state.

It is interesting that both the AFM phase and the *XY* structure of the domain boundary proved to be stable in relation to variations in local correlation parameter *D* over a wide range up to *D* ~ 1.0. Upon further growth of local correlations, however, the domain boundary structure reorganized radically.

The evolution of the antiphase domain boundary upon an increase in parameter *D* is shown in Fig. 1. As *D* grows gradually, the regular structure of the filamentary *XY* phase on the edges of the antiphase domain boundary is broken, while the QP phase emerges and grows to completely displace the filamentary *XY* phase at *D* ~ 1.2, accelerating the boundary transition to QP. With further growth of local correlations *D* > 1.5, the domain boundary broadens and gradually displaces the AFM order. In other words, the AFM → QP phase transition (the large *D* phase) occurs with an increase in the local correlation parameter, due to expansion of the domain boundaries.

It is noteworthy that the QP phase nucleation on the edges of the domain boundary occurs due to the smaller difference between the energies of the QP and *XY* phases there (Fig. 2). In other words, the emergence of the QP phase on the edges is energetically more advantageous than at the center. In Fig. 2, we can see that the difference between the energies of phases in the domain and at the center of the domain boundary is much smaller when the QP phase emerges at the center of the domain boundary (at *D* = 1.2) than with the *XY* phase (*D* = −5). Upon the further growth of *D*, the AFM phase becomes metastable in the domains, and the QP phase becomes stable at the center of the domain boundary.

The study of temperature effects shows that when the temperature in the domain walls of the AFM phase rises at *D* = 1.0, the system moves from the *XY* phase to the QP phase and then to a disordered paramagnetic state. During subsequent cooling to very low temperatures $T = 0.0001$, however, only the QP structure of the domain boundaries is restored; i.e., a temperature hysteresis is observed in the structure of the boundaries.

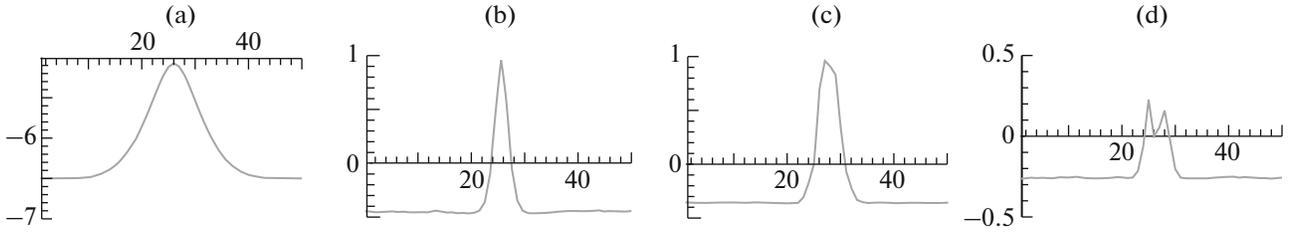

**Fig. 2.** Average distribution across the domain boundary for the local energy at different values of parameter *D*: (a) −5.0, (b) 1.0, (c) 1.1, (d) 1.2. The values along the horizontal axis are presented in terms of the lattice constant; along the vertical axis, the energy is shown in terms of parameter *t*.

## CONCLUSIONS

We studied the effect single-ion anisotropy parameter *D* has on the structure of domain boundaries of the antiferromagnetic phase. Using numerical Monte Carlo modeling on large square lattices with rapid annealing, we observed the formation of a stripe domain structure, in whose antiphase domain boundaries a filamentary *XY* phase formed stably over a wide interval of *D* variations up to positive $D \sim 1$. Upon further growth of local correlations, however, the *XY* phase was broken, and a filamentary QP phase formed in the boundaries separating the domains with antiferromagnetic ordering. Our modeling of temperature effects indicated there was a temperature hysteresis in the structure of the boundaries.

## FUNDING

This work was supported by Program 211 of the Government of the Russian Federation, project no. 02.A03.21.0006; and by the RF Ministry of Science and Higher Education, project nos. 2277 and 5719.